\documentclass[11pt,twoside]{article}
\usepackage{asp2010,natbib}

\resetcounters

\begin{document}

\title{Testing Theory with Dynamical Masses and Orbits of Ultracool
  Binaries}

\author{Trent~J.~Dupuy,$^{1}$ Michael~C.~Liu,$^1$ and Michael~J.~Ireland$^2$
  \affil{$^1$Institute for Astronomy, University of Hawai`i, 2680
    Woodlawn Drive, Honolulu, HI 96822, USA}
  \affil{$^2$Sydney Institute for Astronomy, School of Physics,
    University of Sydney NSW 2006, Australia}}

\begin{abstract}

  Mass is the fundamental parameter that governs the evolution of
  stars, brown dwarfs, and gas-giant planets. Thus, direct mass
  measurements are essential to test the evolutionary and atmospheric
  models that underpin studies of these objects. We present results
  from our program to test models using precise dynamical masses for
  visual binaries based on Keck laser guide star adaptive optics
  astrometric monitoring of a sample of over 30 ultracool ($>{\rm
    M6}$) objects since 2005.  In just the last 2 years, we have more
  than tripled the number of late-M, L, and T dwarf binaries with
  precise dynamical masses.  For most field binaries, based on direct
  measurements of their luminosities and total masses, we find a
  ``temperature problem'' in that evolutionary model radii give
  effective temperatures that are inconsistent with those from model
  atmosphere fitting of observed spectra by 100--300~K.  We also find
  a ``luminosity problem'' for the only binary with an independent age
  determination (from its solar-type primary via
  age--activity--rotation relations).  Evolutionary models
  underpredict the luminosities of HD~130948BC by a factor of
  $\approx$2, implying that model-based substellar mass determinations
  (e.g., for directly imaged planets and cluster IMFs) may be
  systematically overestimating masses.  Finally, we have employed the
  current sample of binary orbits to carry out a novel test of the
  earliest evolutionary stages, by using the distribution of orbital
  eccentricities to distinguish between competing models of brown
  dwarf formation.

\end{abstract}

\section{Introduction}

Binary systems have long been used to probe both the inner workings
and the origins of stars. In multiple systems the laws of gravity can
be used to infer the masses of stars. Mass is the fundamental property
of any object as this largely determines its entire life history, and
thus binaries with direct dynamical mass measurements are important
benchmarks for testing theory. Binaries also record within their
orbits a dynamical imprint of their formation and subsequent
evolution. For a given mass, the total energy of the system is related
solely to the semimajor axis of the orbit ($E = -GM/2a$), while the
angular momentum depends on the orbital eccentricity ($L =
\sqrt{GMa(1-e^2)}$).  Decades before it was known that nucleosynthesis
powers stars, well-determined binary orbits were commonplace
\citep[e.g., $>200$ orbits were published in the compilation
of][]{1918QB821.A3a......} and were, for example, used to argue for a
dynamical constraint on the age of the Galaxy of $\lesssim 10$~Gyr
\citep{ambartsumian37} as opposed to the $\sim$10,000~Gyr age limit
supported by \citet{1935Natur.136..432J}. Thus, the study of binary
orbits has a rich heritage, predating and defining much of modern
astrophysics.

It has only recently been possible to extend such studies below
$0.1~M_{\sun}$, to masses at the bottom of the main sequence and into
the brown dwarf regime.  The field population over this mass range
encompasses spectral types $\gtrsim{\rm M7}$, collectively referred to
as ``ultracool'' owing to their shared low-temperature atmospheric
physics.  Only $\approx$10~years ago were the first large samples of
ultracool dwarfs identified via wide-field surveys such as 2MASS
\citep[e.g.,][]{1999ApJ...522L..65B}, DENIS
\citep[e.g.,][]{1999A&AS..135...41D}, and SDSS
\citep[e.g.,][]{2002astro.ph..4065H}. The underlying binary population
was subsequently uncovered by high-resolution imaging surveys using
\textsl{HST} \citep[e.g.,][]{2003AJ....126.1526B, 2003ApJ...586..512B}
and adaptive optics (AO) from the ground
\citep[e.g.,][]{2002ApJ...567L..53C, 2005astro.ph..8082L}.  In the
first half of the last decade, orbital monitoring of a few select
binaries yielded three dynamical mass measurements: LHS~1070BC
(M8.5+M9.5; \citealp{2001A&A...367..183L}); Gl~569Bab (M8.5+M9;
\citealp{2001ApJ...560..390L}); and 2MASS~J0746$+$2000AB (L0+L1.5;
\citealp{2004A&A...423..341B}). These first direct mass constraints on
ultracool models probed relatively warm temperatures ($\gtrsim2200$~K)
and high masses (only Gl~569Bb is likely to be a brown dwarf;
\citealp{me-latem}).

Over just the last three years the number of dynamical masses
sufficiently precise for meaningful model tests ($\leq30\%$) has
tripled, with components now extending down to masses of
$\approx$30~$M_{\rm Jup}$ and temperatures of $\approx$1000~K. This
rapid expansion in dynamical masses has been driven by high-precision
orbital monitoring programs that utilize the relatively new capability
of laser guide star (LGS) AO to resolve the shortest period binaries
from ground-based telescopes. Our program using Keck LGS AO has been
ongoing since 2005 and has provided five of the six new high-precision
dynamical masses.  All nine ultracool binaries with high-precision
dynamical masses are listed in Table~1.  In the cases where multiple
orbits have been published for a binary, the parameters from the
highest quality orbit determination are given; previous orbit
references are listed in parentheses.

\begin{table}[!ht]
\caption{ Ultracool visual binaries with precise dynamical masses ($\sigma_M/M \leq 30\%$)}
\smallskip
\begin{center}
{\small
\begin{tabular}{lccccc}

\tableline
\noalign{\smallskip}

Name           & Component &        Period           &  Eccentricity          &  Total mass                & Ref.\\
               & Sp. Types &        [days]           &                        &  [$M_{\sun}$]              &     \\

\noalign{\smallskip}
\tableline
\noalign{\smallskip}

Gl~569Bab      & M8.5+M9   &   $864.5\pm1.1$         & $0.316\pm0.005$        &  $0.140^{+0.009}_{-0.008}$ & 1 (2--5) \\
LP~349-25AB    & M7.5+M8   &    $2834\pm15$          & $0.051\pm0.003$        &  $0.120^{+0.008}_{-0.007}$ &  1  (5)  \\
HD~130948BC    &   L4+L4   &    $3760\pm60$          & $0.176\pm0.006$        & $0.109\pm0.002$          & 6 (5,7)  \\
2M0746$+$20AB &   L0+L1.5 &    $4640\pm30$          & $0.487\pm0.003$        &  $0.151\pm0.003$           &  5  (8)  \\
LHS~2397aAB    &   M8+L7   &    $5190\pm40$          & $0.350\pm0.005$        &  $0.146\pm0.014$           &  9  (5)  \\
2M1534$-$29AB &   T5+T5.5 &    $5500^{+800}_{-600}$ & $0.25^{+0.11}_{-0.13}$ &  $0.056\pm0.003$           &  10 (5) \\
LHS~1901AB     & M6.5+M6.5 &    $5880\pm180$         & $0.830\pm0.005$        &  $0.194^{+0.025}_{-0.021}$ &  1  \\
LHS~1070BC     & M8.5+M9.5 &  $6214.7\pm0.4$         & $0.034\pm0.002$        &  $0.157\pm0.009$           &  11 (12) \\
2M2206$-$20AB &   M8+M8   & $12800^{+2200}_{-1800}$ &  $0.25\pm0.08$         &  $0.15^{+0.05}_{-0.03}$    &  13 (5) \\

\noalign{\smallskip}
\tableline
\end{tabular}
}
\end{center}

{\small

{References. ---} (1)~\citet{me-latem}; (2)~\citet{2001ApJ...560..390L}; (3)~\citet{2004astro.ph..7334O}; 
(4)~\citet{2006ApJ...644.1183S}; (5)~\citet{qk10}; (6)~Dupuy~et~al.\ (2011, revised); (7)~\citet{me-130948};
(8)~\citet{2004A&A...423..341B}; (9)~\citet{me-2397a}; (10)~\citet*{2008ApJ...689..436L}; 
(11)~\citet{2008A&A...484..429S}; (12)~\citet{2001A&A...367..183L}; (13)~\citet{me-2206}.

}

\end{table}

\section{Testing Evolutionary and Atmospheric Models}

The growing sample of ultracool dwarf dynamical masses has enabled
some of the strongest tests of theoretical models to date,
particularly in the brown dwarf regime. Because brown dwarfs do not
have a sustained source of internal energy generation, they simply
grow colder and fainter over time. Thus, the sample of field brown
dwarfs spans a wide range of masses at a given luminosity, and the
only means of pinpointing a given brown dwarf's evolutionary status is
to measure its mass or age.  As shown by \citet*{2008ApJ...689..436L},
fundamental properties such as $T_{\rm eff}$ and $\log(g)$ are
$\approx$5$\times$ better constrained (i.e., model tests are
$\approx$5$\times$ stronger) with dynamical mass measurements as
compared to age determinations, which are ordinarily of lower
precision.

One of the major components of our work has been to develop an
analysis method that produces rigorous, quantitative tests of models
based solely on directly measured properties (i.e., mass and
luminosity). This is of fundamental importance, as other previous
approaches were sometimes based on $T_{\rm eff}$ estimates or the
model-predicted color--magnitude diagram, both of which are often
prone to severe systematic biases (e.g., see discussions in
\citealp{2006AJ....131..638G} and Section~4.5 of \citealp{me-latem}).
The discrepancies we have found between observations and models to
date fall into two categories: (1)~problems with the luminosities
predicted by evolutionary models, (2)~problems with temperature
estimates derived from model atmospheres and/or from evolutionary
model radii.

\subsection{``Luminosity Problem''}

Conventional wisdom is that the most robust predictions of substellar
models are bulk properties such as the radius and bolometric
luminosity ($L_{\rm bol}$) since independent models put forth by
different groups produce nearly identical (to within a few percent)
values at ages beyond a few Myr. Consequently, evolutionary models
have become trusted to provide accurate mass estimates in the many
situations when luminosity and age are well constrained, such as
cluster mass functions \citep[e.g.,][]{2007MNRAS.380..712L} and
directly imaged extrasolar planets \citep[e.g.,][]{hr8799}.  However,
until recently no direct measurements of luminosity and mass were
available for any brown dwarfs that also had well determined ages.

The first, and to date only, system that enables a direct test of
substellar luminosity evolution is the L4+L4 binary HD~130948BC
\citep{me-130948}.  Its model-derived age (given its measured total
mass and component luminosities) is significantly younger than the age
of the primary G2V star in this triple system. This is a 2$\sigma$
discrepancy when accounting for the errors in the binary's measured
mass and luminosities and the gyrochronology relation used to age-date
the primary star \citep{mam08-ages}.  If we assume that the stellar
age and its error are correct (i.e., HD~130948A is not simply an
outlier as compared to the cluster rotation sequences), then the
luminosities predicted by evolutionary models for HD~130948B and C are
a factor of $\approx$2 lower than what is observed. In considering
this potential luminosity problem, we will now address the following
questions: (1)~is there additional evidence for models underpredicting
luminosities of substellar objects; and (2)~what physical mechanism
could be responsible for this effect?

\begin{figure}[!ht]
\plottwo{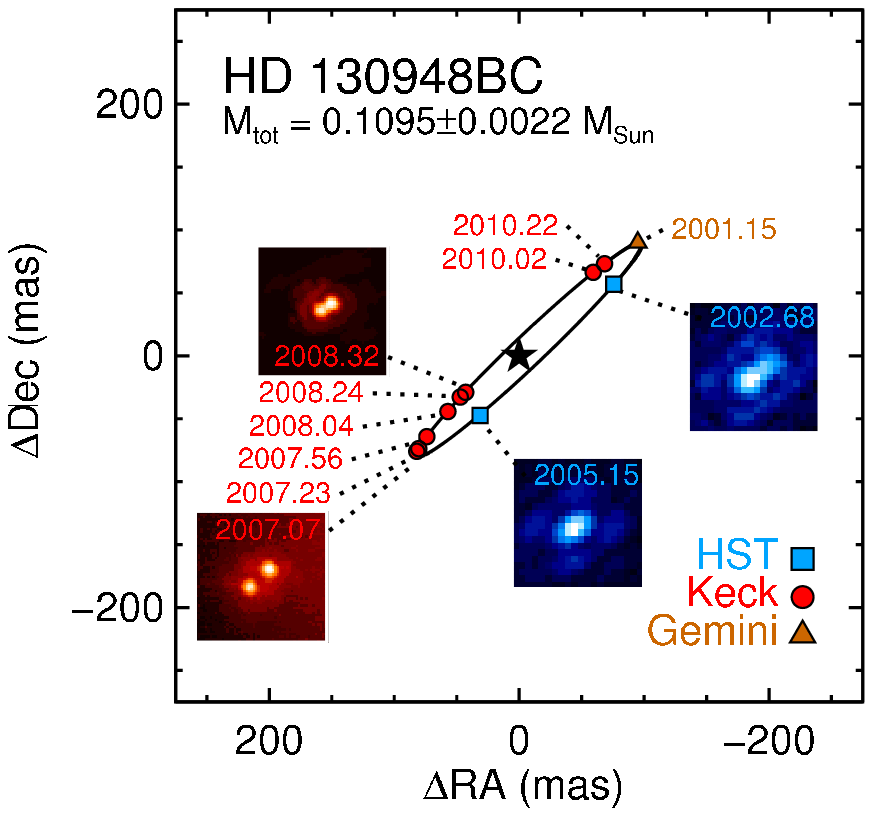}{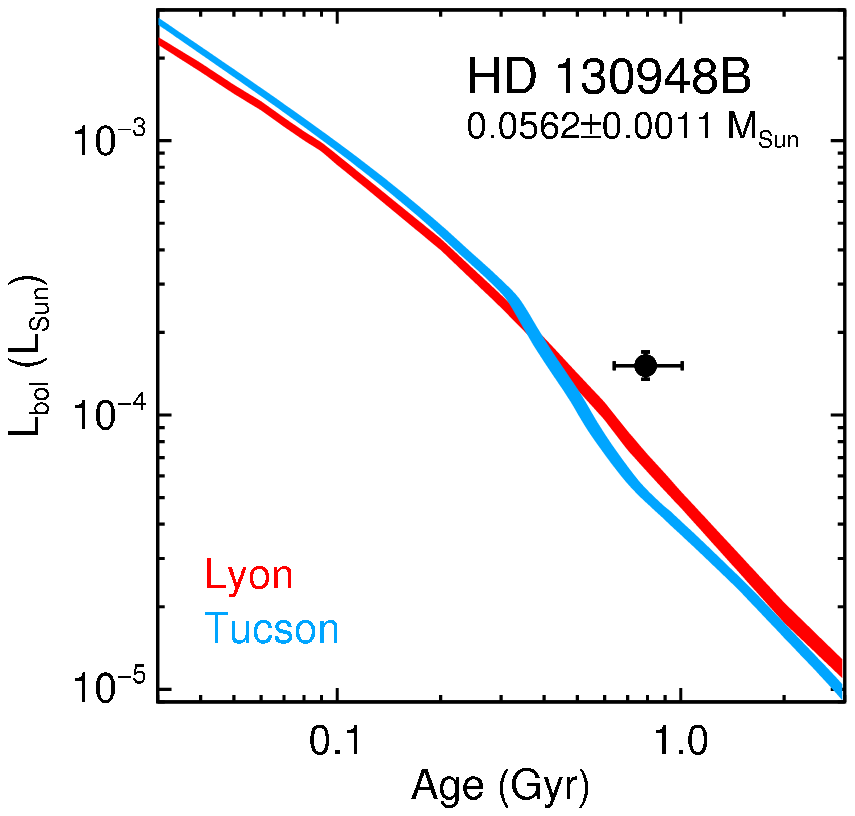}
\caption{\emph{Left:} Relative astrometry of the L4+L4 binary
  HD~130948BC from \textsl{HST}, Gemini, and Keck AO imaging along
  with the best-fit orbit.  \emph{Right:} Given the measured dynamical
  mass, evolutionary models predict the luminosity of the components
  of HD~130948BC as a function of age (colored lines; widths indicate
  the 2.0\% mass error propagated through the models).  The data point
  shows the measured luminosity of HD~130948B at the age determined
  from age--activity--rotation relations for the primary star in this
  triple system ($0.79^{+0.22}_{-0.15}$~Gyr).  Both HD~130948B and
  HD~130948C are a factor of $\approx$2 more luminous than predicted
  by models.}
\end{figure}

\textit{(1)~Is there additional evidence for the luminosity problem?}
If models underpredict luminosities, this would manifest as either
model-derived masses (based on $L_{\rm bol}$ and age) that are
$\approx$25\% too high or model-derived ages (based on mass and
$L_{\rm bol}$) that are too young. The former problem can be tested in
the multiple planet system around HR~8799.  \citet{fm-8799} have shown
that there only two plausible scenarios for the three tightly packed,
massive planets (model-derived masses of 7, 10, and 10~$M_{\rm Jup}$)
to be dynamically stable: (1)~the inner planets must be in 2:1
resonance, and all three planets must have masses lower than predicted
by models by at least 20\%--30\%; or (2)~both the inner pair and outer
pair of planets must be in 2:1 resonances, in which case the masses of
the planets could be as much as a factor of $\sim$2 larger. The
careful construction of the system required in the latter case may
suggest that the first scenario is in fact more plausible, and the
``luminosity problem'' may also be seen here.

Further evidence for the luminosity problem can also come from objects
with model-derived ages that appear to be too young. \citet{me-latem}
have observed this in the M7.5+M8 binary LP~349-25AB for which Lyon
models give an age of $130\pm20$~Myr. This is essentially identical to
the canonical Pleiades age of 125~Myr as determined from the lithium
depletion boundary independent of model luminosity predictions
\citep{1998ApJ...499L.199S}. However, LP~349-25AB appears to be
significantly older than the Pleiades given the strong upper limit on
its \ion{Li}{1} abundance \citep{2009ApJ...705.1416R}, and it also
does not show spectroscopic signatures of youth as is normally seen in
young ($\sim$100~Myr) field dwarfs
\citep[e.g.,][]{2004ApJ...600.1020M, 2007ApJ...657..511A}.

\textit{(2)~What physical mechanisms could explain the luminosity
  problem?} After a few Myr of evolution, model cooling tracks that
adopt very different initial conditions converge to essentially the
same luminosity \citep[e.g.,][]{2002A&A...382..563B}, so accounting
for different formation histories such as in ``cold start'' models
with radiative energy losses due to accretion
\citep[e.g.,][]{2007ApJ...655..541M} will not alter model cooling
tracks sufficiently to explain our observed luminosity discrepancy.
Rather, it is more likely that a process sustained over these objects'
lifetimes, such as convection, is responsible for the
problem. \citet{2010ApJ...713.1249M} have recently constructed
evolutionary models tailored to the components of HD~130948BC in which
the onset of convection is inhibited by very strong magnetic fields
throughout the interior.  This effect can substantially alter the
radius evolution of the components of HD~130948BC and thus explain the
observed discrepancy on the H-R diagram reported by \citet{me-130948}
because their new models yield larger radii, which shifts model tracks
to lower $T_{\rm eff}$.  However, at the same time these models also
\emph{lower} the model-predicted luminosities by $\approx$0.2~dex
($\approx$50\%), increasing the observed luminosity discrepancy.

If the inhibition of the onset of convection has the opposite effect
on the luminosity evolution, then perhaps the efficiency of convection
must be changed. \citet{2007A&A...472L..17C} consider the impact of
strong interior magnetic fields and fast rotation on the convective
energy transport in low-mass stars. They find that these processes
result in lowered convective efficiency, which in turn results in
bloated radii for $\approx$0.3--0.8~$M_{\sun}$ stars, while lower mass
fully convective stars are much less affected. The effect of reduced
convective energy transport for brown dwarfs is not investigated in
detail, but \citet{2007A&A...472L..17C} point out that its main effect
will be to slow their contraction rate. This may effectively ``slow
down the clock'' of substellar evolution, i.e., for a given mass and
age objects would be larger and more luminous than predicted by
conventional evolutionary models that do not include the effects of
convective inhibition due to rotation or strong interior magnetic
fields.

\subsection{``Temperature Problem''}

Model atmospheres offer the possibility of determining the temperature
and surface gravity of an object from a single spectrum.  This
approach is widely used in the study of very low-mass stars, brown
dwarfs, and directly imaged planets.  However, the reliability of such
estimates (and the conclusions drawn from them) depend entirely the
fidelity of the model atmospheres being used.  This can only be
assessed in special cases where independent constraints on the mass,
age, or radius are available.  We have utilized our measured dynamical
masses of ultracool field dwarfs to provide some of the strongest
tests of model atmospheres to date. This is made possible by the fact
that direct mass measurements effectively pin down the evolutionary
status of a field object to a degree of certainty commensurate with
the precision of the measured mass and luminosity (both typically
$\lesssim10\%$). The values of $T_{\rm eff}$ and $\log(g)$ then inferred
from evolutionary models are limited to a very narrow range, typically
exceeding the precision of the model atmosphere grid steps of
50--100~K and 0.5~dex. This evolutionary model-derived $T_{\rm eff}$
is essentially a restatement of the model radius given the measured
luminosity, following the definition of effective temperature, $T_{\rm
  eff} \equiv (L_{\rm bol} / 4\pi R^2 \sigma)^{1/4}$. Therefore, these
binaries provide precise temperature benchmarks that can be compared
to results from model atmosphere fitting.

We have found that temperatures derived from model atmospheres are
typically inconsistent with those from evolutionary models at the
level of $\approx$100--300~K. The amplitude and the sign of this
disagreement vary over the full range of our sample.  For example,
T~dwarf model atmosphere fits give $\approx$100--200~K warmer
temperatures than evolutionary model radii
\citep{2008ApJ...689..436L}, but L~dwarf model atmosphere fits give
$\approx$200--300~K cooler temperatures \citep{me-130948}. The lack of
a systematic offset for all objects indicates that the disagreement is
not simply due to a uniform error in predicted radii at all masses and
spectral types, but without directly measured radii, which would
provide $T_{\rm eff}$\ directly, it is impossible to distinguish whether
evolutionary or atmospheric models are responsible for the observed
discrepancies.

If we restrict our comparisons to a narrow range of temperatures, we
can essentially remove one variable from the problem and focus on how
the $T_{\rm eff}$ discrepancies vary with mass and age. To date, the
largest subset of mass benchmarks at any given temperature are the
late-M dwarfs (Table~1). We obtained integrated-light near-infrared
spectroscopy of four nearly equal-flux late-M binaries with component
mass determinations ranging from $\approx$0.06--0.10~$M_{\sun}$ (i.e.,
young brown dwarfs to old stars) and with a range of chromospheric
activity levels \citep{me-latem}. After fitting these spectra with
four independent model atmosphere grids, we found that all best-fit
temperatures were systematically $\approx$250~K warmer than the
$T_{\rm eff}$ values derived from evolutionary models
(Figure~2). Thus, the observed $T_{\rm eff}$ discrepancy is the same
for objects with very similar spectra but with very different masses,
ages, and activity levels, indicating that the model atmospheres are
largely responsible for the observed $T_{\rm eff}$ discrepancy.

\begin{figure}[!t]
\plotone{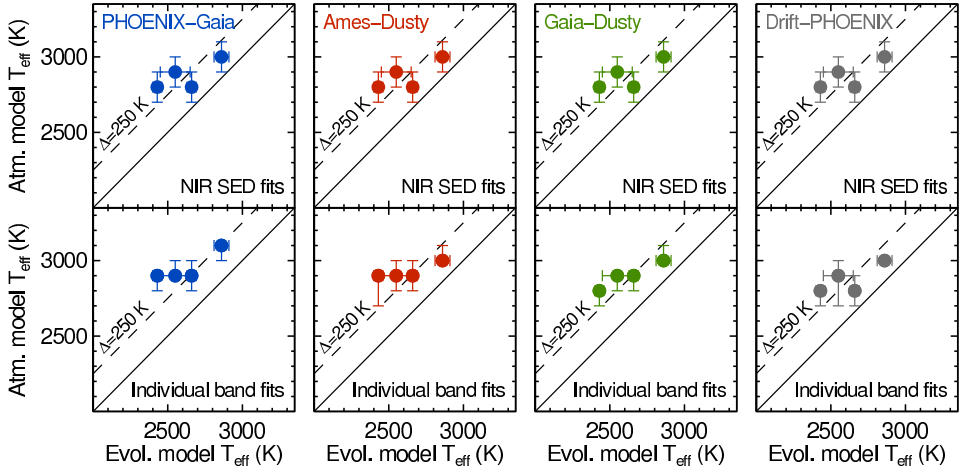}
\caption{Late-M dwarf effective temperatures determined from model
  atmosphere fitting compared to Lyon Dusty evolutionary model-derived
  $T_{\rm eff}$, (i.e., from measured total mass and individual luminosities
  \citealp{me-latem}).  Four independent model atmosphere grids were
  tested (left to right) and two different methods of fitting
  atmospheres: fitting the full 0.95--2.42~$\mu$m near-infrared
  spectrum (top panels), and fitting the $Y$, $J$, $H$, and $K$ bands
  individually (bottom panels).  In all cases, an offset of 250~K
  between the two classes of models is observed (dashed line).  This
  implies that either evolutionary model estimates are too cool (i.e.,
  radii too large by 15\%--20\%) or that atmospheric model estimates
  are too warm by 250~K.  The latter is more likely given that this
  sample spans a wide range of masses, ages, and activity levels, but
  a narrow range of $T_{\rm eff}$.}
\end{figure}

\section{Testing Formation Models with Eccentricities}

Some properties of very low-mass binaries, such as orbital semimajor
axes, mass ratios, and multiplicity fractions, can be studied in a
statistical fashion using a large sample without any knowledge of
orbital motion \citep[e.g.,][]{2007ApJ...668..492A}, and such studies
have provided very important constraints on formation models at very
low masses.  However, only well-determined orbits can provide
eccentricities ($e$), which are directly related to the angular
momentum of the formation process.  In fact, the only two theoretical
simulations that report eccentricities of very low-mass binaries
predict very different types of orbits: (1)~the cluster formation
model of \citet{2009MNRAS.392..590B} yields a total of 16 binaries
with typically modest eccentricities, and (2)~the disk gravitational
instability model of \citet{2009MNRAS.392..413S} produces 12 binaries
with typically high eccentricities (Figure~3).

\begin{figure}[!t]
\plottwo{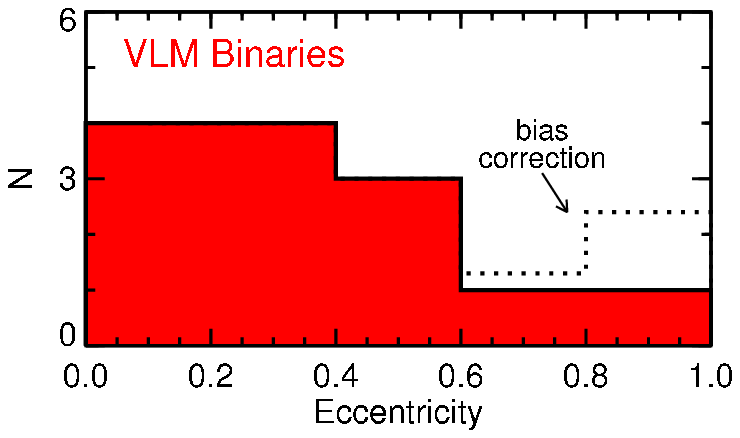}{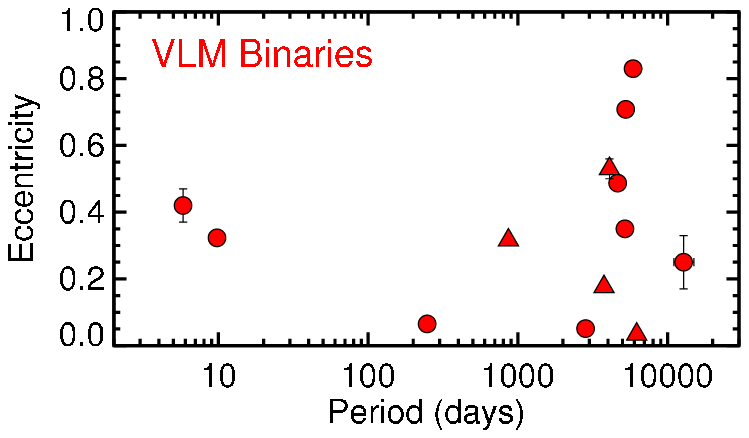}
\plottwo{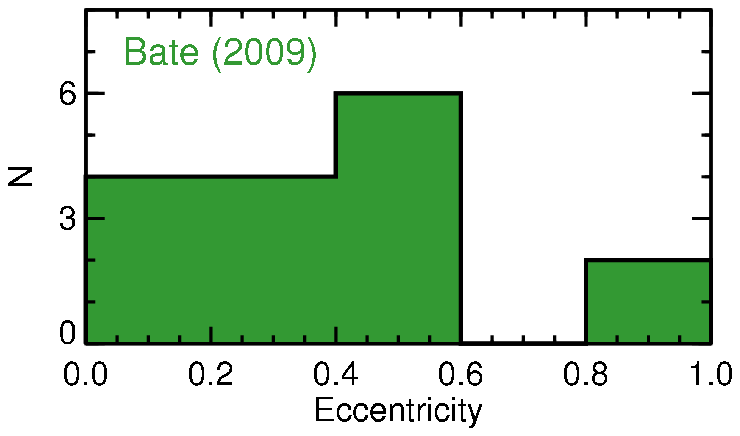}{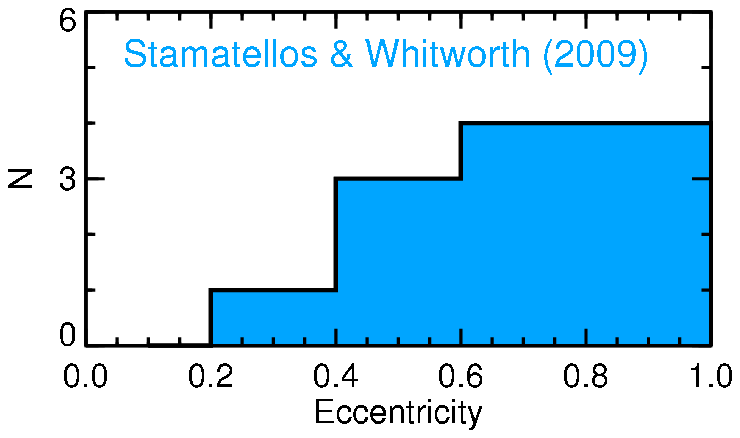}
\caption{Eccentricity distribution of all very low-mass binaries with
  well-determined orbits, including both visual and spectroscopic
  binaries (top panels). The observed distribution is very similar to
  that predicted by the cluster formation model of
  \citet[][bottom left]{2009MNRAS.392..590B} but is inconsistent with the
  much higher eccentricities predicted by the gravitational instability
  model of \citet[][bottom right]{2009MNRAS.392..413S}, even after
  applying a bias correction to the observed distribution.}
\end{figure}

The sample of very low-mass binaries with well-determined orbits is
now of comparable size to the binaries produced in theoretical
simulations, and thus it can be used to try to discriminate between
the very different model predictions. Figure~3 shows the eccentricity
distribution for all very low-mass binaries with eccentricity
uncertainties less than 0.1. This excludes the visual binary
2MASS~J1534$-$2952AB (Table~1) and includes three spectroscopic
binaries: PPl~15 ($e = 0.42\pm0.05$; \citealp{1999AJ....118.2460B}),
2MASS~J0535$-$0546 ($e = 0.323\pm0.006$;
\citealp{2006Natur.440..311S}), and 2MASS~J0320$-$0446 ($e =
0.065\pm0.016$; \citealp{2008ApJ...678L.125B}). This sample populates
a broad range in eccentricity from nearly circular to highly eccentric
($e$ = 0.03--0.83) and shows a strong preference for modest
eccentricities ($\bar{e} = 0.35$, median of 0.32).  The observed
distribution of eccentricities is well-matched to that predicted by
the cluster formation model of \citet{2009MNRAS.392..590B}, although
we note that this simulation is not in full agreement with
observations as it produces much wider binaries than are observed in
the field. In contrast, the gravitational instability model of
\citet{2009MNRAS.392..413S} produces too many high-$e$ binaries, with
an absence of \emph{any} modest eccentricity ($e<0.3$) orbits, which
is highly inconsistent with the observed distribution.

\end{document}